\RequirePackage[2020-02-02]{latexrelease}
\documentclass[amscls]{revtex4}
\usepackage{amsmath}
\newcommand{\be}{\begin{equation}}
	\newcommand{\ee}{\end{equation}}
\newcommand{\ba}{\begin{eqnarray}}
	\newcommand{\ea}{\end{eqnarray}}
\begin{document}

	\title{Affine Connection and Quantum Theory} 
	\author{Kaushik Ghosh\footnote{E-mail kaushikhit10@gmail.com}}
	\affiliation{Vivekananda College (University of Calcutta), 
		269, Diamond Harbour Road, Kolkata - 700063, India}

	\maketitle
	
	\section*{Abstract}

	In a few recent manuscripts, we used the affine connection to introduce two massless scalar fields in the Einstein-Palatini action. These fields lead to non-metricity. In this article, we will discuss the significance of these fields in inflation and dark energy. We will construct a Lagrangian formalism to include these scalar fields in a theory of gravity coupled with ordinary matter and radiation. We will find that these fields need not to be included in the actions of interacting gauge theories coupled with conserved fermionic vector currents as a part of the connection. The same remains valid for ordinary scalar fields. We can couple the connection-scalars with ordinary matter by adding suitable interaction terms. In this context, we will find that Stokes' theorem leads us to include the right-handed neutrinos in the electroweak theory in curved spacetime even with the Levi-Civita connection. This is required to obtain consistent equations of motion and anomaly-free conserved vector current for the neutrinos. Axial vector currents for different Dirac fields may remain anomalous. The right-handed neutrinos can be useful to explain neutrino oscillation and dark matter. We will also discuss the possibility of introducing massless finite integer spin particles using second rank symmetric traceless tensors with reference to  the corresponding little group in flat spacetime. We will show that we can use massless $(A,A)$ type fields in Minkowski space to introduce massless finite integer spin particles.

	\vspace{0.5cm}

	
	Key-words: quantum fields in curved spaces, anomaly, right handed neutrino, inflation, dark energy

	\vspace{0.5cm}

	MSC: 53Z05, 49S05, 83D05, 81T20, 81T50

	\vspace{0.5cm}

	PACS: 02.40.-k, 98.80.Qc, 98.80.Cq, 95.36.+x, 04.50.Kd

	\section*{I. Introduction}

A lot of work has been done during the last few decades on two periods of cosmic accelerations: inflation and dark energy [1,2,3,4,5,6]. Inflation is driven by the scalar field inflation. We also have multi-scalar inflation and higher spin driven inflation models [7,8,9]. Dark energy is not yet detected directly. It is not localized like ordinary matter and is assumed to have negative pressure to explain the present cosmic acceleration. We also need dark matter which is observed by its gravitational effects [6,10,11,12]. Data from the Cosmic Microwave Background Radiation (CMBR) and supernovae surveys predict that the energy composition of the universe consists of about $20 \%$  dark matter, about $70 \%$ dark energy and the rest ordinary matter like hadrons and radiation. The first candidate for dark energy is the cosmological constant $\Lambda$ [6,13]. It is of order $(10^{-3}eV)^4$ and is far below from the value predicted by quantum gravity [14]. This led to construct alternate theories. There are two principal approaches. The first adds stress-tensors with negative pressure in the source part of the Einstein equation. Quintessence, k-essence and perfect fluid models are the most well-known among various theories [15,16,17]. These fields are expected to originate as matter fields. There is the possibility that they can come with a negative kinetic term [5,6]. Corresponding fields are known as ghost fields.
Despite a lot of efforts, the origins of inflation, dark energy and dark matter are yet to completely fit within an extended version of the standard model. The other approach modifies the Einstein-Hilbert action. Examples are $f(R)$ gravity [18,19], scalar-tensor theories [6,20] and braneworld models [21,22]. These are geometric theory of dark energy [20]. The experimental success of Einstein's theory of gravity, linear in the scalar curvature, in explaining many 
astrophysical observations led us to search for additional fields that can be introduced by using a quantum theory of gravity that is linear in the scalar curvature at the classical level.

In a few recent manuscripts [23,24], we have demonstrated that a quantum theory of gravity can introduce additional fields besides metric that can be useful to explain inflation and dark energy. Here, we are considering the free theory. The natural candidates are the affine connections used to define covariant derivatives in a curved spacetime [25,26,27,28,29,30]. We found it is more appropriate to use affine connection more general than the metric compatible connections in a quantum theory of gravity [23,24]. This is related to the consistency of canonical commutators and various metric-metric commutators used to quantize the theory. A theory involving curvature is not primarily a theory of metric but in general is a theory of non-trivial connection [26]. Hence, it is not bizarre that we can have non-metricity [31] in a quantum theory of gravity. Regarding variational principle, the simplest theory to introduce connections more general than the Levi-Civita connection is to consider the Einstein-Palatini action [28]. Here, the connection is expressed as:
$\Theta^{\alpha}_{~\mu \nu} = \Gamma^{\alpha}_{~\mu \nu} + C^{\alpha}_{~\mu \nu}$,
where $\Gamma^{\alpha}_{~\mu \nu}$ is the metric compatible Levi-Civita connection and $C^{\alpha}_{~\mu \nu}$ is a third rank tensor field that can introduce non-metricity. The Lagrangian density is given by $\sqrt{-g} R'$, where $R'$ is the scalar curvature evaluated using the connection $\Theta^{\alpha}_{~\mu \nu}$. In [23,24], we have used a potential-like formalism to express $C^{\alpha}_{~\mu \nu}$ as various covariant derivatives of lower rank tensor fields. This is similar to the Levi-Civita connection which is constructed from metric. We will find in Sect.III that scalar fields are the most suitable candidates for this purpose. This is due to the compatibility of the corresponding Einstein equation with the contracted Bianchi identity in $\Gamma^{\alpha}_{~\mu \nu}$. Unless stated otherwise, in this article, contracted Bianchi identity will mean twice contracted Bianchi identity. We have previously introduced two scalar fields that obey homogeneous wave equation in the free theory and can be non-localized. These scalar fields can be related to two traces of $C^{\alpha}_{~\mu \nu}$, one \textit{w.r.t} $(\alpha,\mu)$ and the other \textit{w.r.t} $(\mu,\nu)$. They provide positive and negative stress-tensors in Einstein's equation respectively, and will be useful for constructing new non-trivial vacuum solutions [29]. Here, we have a scalar-tensor theory directly related to quantum gravity, that could explain both inflation and dark energy provided we modify the action by adding suitable potential terms [6]. 
The scalar fields mentioned above give non-metricity although their effects are geometrically different. They are similar to the cosmological constant $\Lambda$ that gives deviation from the Newtonian theory of gravity in the weak field slow motion limit [29]. In the present article, we will elaborate on this in the next section. We will mostly consider affine connection symmetric in the lower indices that gives symmetric Ricci tensor. In Sect.II, we will first discuss the scalar field $\omega$ that can be related to an affine connection compatible with a metric conformal to the actual solution. This introduces a massless scalar field similar to dilaton. We will also introduce a massless scalar field $\psi$ that is more significant as far as non-metricity is concerned. $\psi$ contributes a negative stress tensor in Einstein's equation and can be useful to explain dark energy.

In Sect.III, we will discuss the possible coupling between $\omega$, $\psi$ and matter fields including radiation that 
can arise due to the presence of $C^{\alpha}_{~\mu \nu} (\omega, \psi)$ in the covariant derivative.
Matter fields are various representations of $SL(2,c)$. In this context, a careful application of the Stokes' theorem reveals that we may need to introduce the right-handed neutrinos as free fields in the electroweak theory in curved spacetime. This remains valid even when we use only the Levi-Civita connection and is required to obtain consistent equations of motion and anomaly-free conserved vector current for the neutrinos. The right-handed neutrinos can be important in dark matter research. There are other advantages related to neutrino oscillations to be mentioned later.  Stokes' theorem also indicates that the axial vector currents for various Dirac fields, including that of neutrinos, could become anomalous in curved spacetime as well as in flat spacetime when we use curvilinear coordinates. This is not rare in quantum field theory in flat spacetime. We will demonstrate that in a mathematically consistent Lagrangian description that preserves the principle of general covariance and the principle of equivalence, it is sufficient to use ${\Gamma^{\alpha}_{~\mu \nu}}$ as the connection to construct actions and conserved stress tensors of ordinary matter and radiation fields. We will find that for fields relevant to the standard model with the addition of the right handed neutrinos, $C^{\alpha}_{~\mu \nu} (\omega, \psi)$ can at most contribute boundary terms to the matter and radiation field actions and do not appear in the equations of motion. Thus, we may not include $C^{\alpha}_{~\mu \nu} (\omega, \psi)$ in the actions and stress tensors of ordinary matter and radiation. The resulting equation is consistent with the contracted Bianchi identity in $\Gamma^{\alpha}_{~\mu \nu}$,
and merely contains an additional stress-tensor in the right hand side of Einstein's equation coming from $C^{\alpha}_{~\mu \nu} (\omega, \psi)$ present in the curvature scalar. This agrees partly with experimental observations regarding dark energy which does not couple with ordinary matter and radiation. We will find in Sect.IV that this may also be required to construct a stable theory [6].  Deviation from Einstein's gravity with $\Gamma^{\alpha}_{~\mu \nu}$, due to dark energy, is small. This is apparent from the smallness of the cosmological constant in the $\Lambda$CDM theory. A similar situation presently remains valid for the stress-tensor associated with $\omega$ and $\psi$, if we try to explain dark energy using these fields. This explains why the Levi-Civita connection and Einstein's gravity describe the observable universe to a very good approximation even if $\omega$ and $\psi$ are present. Non-metricity produced by $\omega$ and $\psi$ is also small.

In Sect.IV, we will discuss possible modifications of Einstein's theory of gravity in the presence of $\omega$ and $\psi$. The equations for $\omega$ and $\psi$ are coupled through derivative terms. We will find that we can separate the equations by introducing two scalar fields, $\Phi$ and $\Psi$ that are linear combinations of $\omega$ and $\psi$. 
The resulting action is a new form of scalar-tensor theory with two scalar fields. $\Phi$ contributes a positive stress tensor to Einstein's equation. When required, we can supplement the Einstein-Palatini action by $\Phi$ dependent potential term. We can couple $\Phi$ with ordinary matter by introducing suitable interaction term. $\Phi$ gives finite non-metricity and we can construct a quantum theory where only $\Phi$ is present. Later, we will consider theories with more than one such field. $\Psi$ contributes a negative stress-tensor. A scalar matter with negative kinetic term in Einstein's equation is known as a phantom/ghost scalar [6]. We find that a phantom scalar of geometric origin can be present in quantum gravity. The negative kinetic term of $\Psi$ can lead to instability if coupled to ordinary matter. We can add $\Psi$ dependent potential term to get stable theories [6]. We will mention a few words on the stability of the theory in this section. Another way to generalize Einstein's theory is to include $\Phi$, $\Psi$ and the cosmological constant $\Lambda$. This could lead to a possible geometric theory of inflation together with dynamic and spatially varying dark energy.

In section:V, we will first discuss the possibility of introducing additional scalar fields and higher spin fields. We will find that we need to modify the Einstein-Hilbert-Palatini action if we want to introduce higher spin fields using the affine connection. In flat spacetime, second rank symmetric traceless tensors give $(A,A)$ type representations of higher spin fields like $j = 1,2$. The most general candidate to give the stabilizer (little group) for massless fields in Minkowski space is taken to be: $W = ISO(2)$ [32]. However, this group is not represented by normal matrices [33] when acting on four vectors/tensors. We will find that we have to use a one parameter subgroup of $SO(3)$ as the little group for massless higher spin fields in Minkowski space. This allows us to construct a $(A,A)$ type quantum field theory for massless vector fields using symmetric traceless derivatives. Such derivatives are not gauge invariant and cannot couple with matter fields in a theory that preserves local internal gauge invariance. They can be useful for describing dark energy and dark matter. The massless higher spin particles can also be useful to construct higher spin driven inflation theories [9,16] in the absence of spontaneous symmetry breaking. Such a theory need not to be a gauge theory.

	\subsection*{II. Scalar Fields from Affine Connection}

The simplest theory to introduce connections more general than the Levi-Civita connection is to consider the Einstein-Palatini action [29]. Here, the connection is expressed as:
$\Theta^{\alpha}_{~\mu \nu} = \Gamma^{\alpha}_{~\mu \nu} + C^{\alpha}_{~\mu \nu}$,
where $\Gamma^{\alpha}_{~\mu \nu}$ is the metric compatible Levi-Civita connection and $C^{\alpha}_{~\mu \nu}$ is a third rank tensor field that can introduce non-metricity. In this section we will consider $C^{\alpha}_{~\mu \nu}$ that is symmetric in the lower indices. The curvature and Ricci tensors are given by the following expressions:

\ba
{{\nabla'}_{\mu}}{A_{\nu}} & = & {\partial}_{\mu}{A_{\nu}} -  
{{\Theta}^{\alpha}_{~\mu \nu}}{A_{\alpha}} \\ \nonumber
({{\nabla'_{\mu}}{\nabla'_{\nu}}} - {{\nabla'_{\nu}}{\nabla'_{\mu}}}){A^{\alpha}_{~\beta}}
& = & - {{R'}_{\mu \nu \kappa}^{~~~~\alpha}}{A^{\kappa}_{~\beta}} +
{{R'}_{\mu \nu \beta}^{~~~~\kappa}}{A^{\alpha}_{~\kappa}} \\ \nonumber
{R'}_{\mu \nu \alpha}^{~~~~\kappa} & = & {{R}_{\mu \nu \alpha}^{~~~~\kappa}} + 
2{{\nabla}_{[\nu}}{C^{\kappa}_{~\mu] \alpha}}
+ 2 [{C^{\lambda}_{~[\mu |\alpha|}} {C^{\kappa}_{~\nu |\lambda|]}}]\\ \nonumber
{R'}_{\mu \alpha} & = & {{R}_{\mu \alpha}} + 
2{{\nabla}_{[\kappa}}{C^{\kappa}_{~\mu] \alpha}}
+ 2 [{C^{\lambda}_{~[\mu |\alpha|}} {C^{\kappa}_{~\kappa |\lambda|]}}]\\ \nonumber
R' & = & {g^{\mu \alpha}} [{{R}_{\mu \alpha}} + 
2{{\nabla}_{[\kappa}}{C^{\kappa}_{~\mu] \alpha}}
+ 2 [{C^{\lambda}_{~[\mu |\alpha|}} {C^{\kappa}_{~\kappa |\lambda|]}}]
\ea	

\noindent{where ${{\nabla'}_{\mu}}$ is the covariant derivative evaluated using the complete connection ${{\Theta}^{\alpha}_{~\mu \nu}}$, ${\nabla_{\mu}}$ is the covariant derivative evaluated using 
${\Gamma^{\alpha}_{~\mu \nu}}$, ${{R}_{\mu \nu \alpha}^{~~~~\kappa}}$ is the curvature tensor for  ${\Gamma^{\alpha}_{~\mu \nu}}$ and ${{R}_{\mu \alpha}}$ is the Ricci tensor associated with the Levi-Civita connection. The Lagrangian density is given by $\sqrt{-g} R'$, where $R'$ is the scalar curvature evaluated using the connection $\Theta^{\alpha}_{~\mu \nu}$. 
When we apply the principle of least action to obtain the classical configurations, we obtain algebraic constraints on 
$C^{\alpha}_{~\mu \nu}$ that lead to the null solution [29]. In [23,24,25], we have used a potential-like formalism to express $C^{\alpha}_{~\mu \nu}$ as various covariant derivatives of lower rank tensor fields. This is similar to the Levi-Civita connection which are constructed from metric. We will find in the following section after Eq.(6) at point (1) that scalar fields are the most suitable candidates for this purpose. This is due to the compatibility of the corresponding Einstein equation with the contracted Bianchi identity in $\Gamma^{\alpha}_{~\mu \nu}$. Here, we express $C^{\alpha}_{~\mu \nu}$ as:}

\ba
{{\Theta}^{\alpha}_{~\mu \nu}} & = & {\Gamma^{\alpha}_{~\mu \nu}} + {{C}^{\alpha}_{~\mu \nu}} \\ \nonumber
{{C}^{\alpha}_{~\mu \nu}} & = & {{\delta}^{\alpha}_{~ (\mu}}{{\nabla}_{\nu)}}{\omega}
- {1 \over 2}{g_{\mu \nu}}{{{\nabla}^{\alpha}}{\omega}} + 
{g_{\mu \nu}}{{\nabla}^{\alpha} {\psi}}.
\ea

\noindent{The reason for this choice is that the first two terms of ${{C}^{\alpha}_{~\mu \nu}}$ gives the contribution of the conformal scalar field when we construct the connection to be compatible with:}

\be
{\tilde{g}_{\mu \nu}} = (1 + \phi(x)){g_{\mu \nu}}, ~~ \omega = ln(1 + \phi).
\ee

\noindent{We can associate $\omega / \phi$ and $\psi$ with two traces of $C^{\alpha}_{~\mu \nu}$, one in the mixed indices and the other in the lower indices. We find that: $C^{\alpha}_{~\alpha \nu} = {\nabla}_{\nu} (2\omega - \psi)$ and $C^{\alpha \mu}_{~~~\mu} = {\nabla}^{\alpha} (4\psi - \omega)$. We can add additional scalar fields depending on observations. We will discuss this in Sect. V. In the present section we will confine our attention to $\omega$ and $\psi$ only. We will briefly discuss the geometric significance of $\omega$ and $\psi$ in Sect. V [25]. Eq.(2) gives us a symmetric Ricci tensor. The Einstein-Palatini action is now given by the following expression:}

	\be
	S = {\int}{\sqrt{-g}}{R'}{\bf e} + {\sum}{\kappa_{M}}{S_M}({\chi_a}, g_{\mu \nu})
	= {\int}{\sqrt{-g}}{\mathcal{R'}}{\bf e} + {\sum}{\kappa_{M}}{S_M}({\chi_a}, g_{\mu \nu})
	\ee
	
	\noindent{where ${\bf e}$ is the coordinate volume element ${dx^0} \wedge {dx^1} \wedge {dx^2} \wedge {dx^3}$ [29], $\chi_a$ represents collection of ordinary matter and radiation and $S_{M}$ is the corresponding action. In the last expression we have removed total divergence terms coming from the second term, $2{{\nabla}_{[\kappa}}{C^{\kappa}_{~\mu] \alpha}}$, present in the expression of ${R'}_{\mu \alpha}$ given by Eq.(1). 
	This requires fixing the normal derivatives of $\omega$ and $\psi$ on the boundary. Later, we will also require to fix $\omega$ and $\psi$ on the boundary to obtain corresponding equations of motion. Thus, we need to remove ambiguities (if any in curved spacetime) present in solutions with either the Dirichlet boundary conditions or the Neumann boundary conditions. It is possible to use the Cauchy boundary conditions with open boundaries in flat spacetime. When possible, we can use periodic boundary conditions. We can also modify the action by introducing a suitable boundary term with the boundary integrand given by: $- C^{\alpha ~\mu}_{~\mu} + C^{\mu ~\alpha}_{~\mu}$, [25]. This is similar to that done for metric [29] and Chern-Simons theory. Thus, we introduce the modified Ricci tensor ${\mathcal{R'}_{\mu \nu}}$ and modified curvature scalar $\mathcal{R'}$ given by the following expressions:}

	\ba
	{\mathcal{R'}} & = & {g^{\mu \alpha}}{\mathcal{R'}_{\mu \alpha}}
	= {g^{\mu \alpha}} \left\lbrace  {{R}_{\mu \alpha}} 
	+ 2 [{C^{\lambda}_{~[\mu |\alpha|}} {C^{\kappa}_{~\kappa |\lambda|]}}] \right\rbrace  \\ \nonumber
	& = & {R} - {3 \over 2}{({\nabla}{\omega})^{2}} 
	+ 3{({\nabla}{\psi})^{2}} + 3 [({{\nabla}_{\kappa}{\omega}})({{\nabla}^{\kappa}{\psi}})]
	\ea

	\noindent{where $R_{\mu \alpha}$ and $R$ are the Ricci tensor and curvature scalar associated with the Levi-Civita connection respectively. We obtain the field equations by considering
		various variational derivatives. Here, we assume that ${C^{\alpha}_{~\mu \nu}} (\omega,\psi)$ is not present in $S_M$. This is consistent with observations. We will illustrate this later in this section. We have the following modification of the \textit{r.h.s} of Einstein's equation [23,24,25]:}

	\ba
	{{G}_{\mu \alpha}} & = & {8 \pi}[{{P}_{\mu \alpha}(\chi_a)} + {3 \over {16 \pi}}{{P}_{\mu \alpha}(\omega)} - {3 \over {8 \pi}}{{P}_{\mu \alpha}(\psi)} - 
	{3 \over {8 \pi}}{{P}_{\mu \alpha}(\omega,\psi)}] \\ \nonumber
	{{P}_{\mu \alpha}(\omega, \psi)} & = & 
	({{\nabla}_{( \mu}}{\psi})({{\nabla}_{\alpha )}}{\omega})- {1 \over 2}{g_{\mu\alpha}}({{\nabla}_{\kappa}{\psi})({{\nabla}^{\kappa}}{\omega}}) \\ \nonumber
	& = & {1\over {(1 + \phi)}}
	[({{\nabla}_{( \mu}}{\psi})({{\nabla}_{\alpha )}}{\phi})
	- {1 \over 2}{g_{\mu\alpha}}({{\nabla}_{\kappa}{\psi})({{\nabla}^{\kappa}}{\phi}})] \\ \nonumber
	{{{\nabla}_{\kappa}}{{\nabla}^{\kappa}}{\omega}} & = & 
	{1 \over {1 + \phi}}{{{\nabla}_{\kappa}}{{\nabla}^{\kappa}}{\phi}} -
	{({1 \over {1 + \phi}})^2}{({\nabla}{\phi})^{2}} = 0 \\ \nonumber
	{{{\nabla}_{\kappa}}{{\nabla}^{\kappa}}{\psi}} & = & 0 
	\ea

	\noindent{where ${{P}_{\mu \alpha}(\chi_a)}$ is the stress tensor of ordinary matter and radiation. $\omega$ and $\psi$ obey the massless Klien-Gordan equation that have non-trivial solutions for most spacetimes. The above equations are consistent with the contracted Bianchi identity associated with ${\Gamma^{\alpha}_{~\mu \nu}}$: ${{\nabla}^{\alpha}}{{G}_{\mu \alpha}} = 0$. We will discuss this aspect further in the following section. ${{P}_{\mu \alpha}}(\omega)$ and ${{P}_{\mu \alpha}}(\psi)$ are the stress tensors of ordinary massless scalar fields. $\psi$ contributes a negative stress tensor to the Einstein equation. $\omega$ and $\psi$ together with suitable additional interaction terms give an alternate way to explain inflation and dark energy [6,35]. Further significance of negative stress tensor can be found in [36,37]. We find that the coupling of $\omega$ with $\psi$ in ${\mathcal{R'}}$ gives another contribution to the source stress tensor. We will later show that the equations can be separated in the scalar fields. ${{C}^{\alpha}_{~\mu \nu}}$ in Eq.(2) is now given by the solutions of the above equations. ${{P}_{\mu \alpha}(\chi_a)}$ will be multiplied by appropriate factors when we transform to non-geometrized ordinary units.}

	\subsection*{III. Matter Fields in Presence of Non-metricity Scalar Fields}

In this section, we will discuss the possible coupling between the scalars ($\omega$, $\psi$) and ordinary matter fields, including radiation. We have to consider a few points when we try to construct a Lagrangian description of gravity coupled with matter fields with a connection more general than the Levi-Civita connection:

{~}

	(\textbf{1}) \textbf{Euler-Lagrange's equations, Bianchi identity and its contracted forms}:

	The first is regarding the Euler-Lagrange's equations, Bianchi identity and its contracted forms. We do not have any problem with these identities for ${R'}_{\mu \nu \alpha}^{~~~~\kappa}$ given by Eq.(1) when derivatives are taken \textit{w.r.t} the total connection ${\Theta^{\alpha}_{~\mu \nu}}$. $\omega$ and $\psi$ in Eq.(2) can be any pair of regular fields. We can in general have problems when we do away with the total divergence term in $R'$ given by $2{g^{\mu \alpha}}{{\nabla}_{[\kappa}}{C^{\kappa}_{~\mu] \alpha}}$ and apply the principle of least action to $S$ given by Eq.(4). It is required to have: ${{\nabla}^{\alpha}}{{G}_{\mu \alpha}} = 0$, the contracted Bianchi identity \textit{w.r.t} ${\Gamma^{\alpha}_{~\mu \nu}}$, when we express Einstein's equation in the form given by Eq.(6) with the fields associated with ${C^{\alpha}_{~\mu \nu}}$ contributing source stress tensor terms in the right hand side. This remains valid for $\omega$ and $\psi$ individually.  The only scalar quadratic in the first order derivative of a scalar field is the square of the norm of the covariant derivative and the corresponding term in $\sqrt{-g}{\mathcal{R'}}$ coincides with the Lagrangian density of a massless scalar field apart from numerical factor. This leads to the familiar conserved stress tensor of a massless scalar field with a numerical factor in the \textit{r.h.s} of Einstein's equation when we express it in the form given by Eq.(6). The contracted Bianchi identity \textit{w.r.t} ${\Gamma^{\alpha}_{~\mu \nu}}$ also remains valid for Eq.(6) where we have both $\omega$ and $\psi$ with derivative coupling. We will show this also in the next section by diagonalizing the second order expression in ${\nabla_\mu}{\omega}$ and ${\nabla_\nu}{\psi}$ present in ${\mathcal{R'}}$ given by Eq.(5). This observation allows us to introduce scalar fields in ${C^{\alpha}_{~\mu \nu}}$. We can also include various potential terms to be discussed later. Similar situation may not remain valid when we include higher spin fields in ${C^{\alpha}_{~\mu \nu}}$.

	{~}

	(\textbf{2}) \textbf{Application of Stokes' theorem and conservation laws}:

	The second is the possibility of removing the total divergence term: ${\int}{\sqrt{-g}}{g^{\mu \alpha}}{{\nabla}_{[\kappa}}{C^{\kappa}_{~\mu] \alpha}}{d^4}x$ by applying the Stokes' theorem. This is possible when 
	we can express this term as: ${\int}{d^4}x {{\nabla}_{\mu}}({{\sqrt{-g}} V^{\mu}})$, where $V^{\mu}$ is a contravariant vector field [27,29]. This happens when the connection in ${{\nabla}_{\kappa}}$ is ${\Gamma^{\alpha}_{~\mu \nu}}$ [27,29]. We can add a ${C^{\alpha}_{~\mu \nu}}$ only when the total connection is metric compatible, leading to: ${C^{\alpha}_{~\alpha \nu}} = 0$, [27]. The same conditions remain valid when ${C^{\alpha}_{~\mu \nu}}$ is not symmetric in the lower indices. Thus, we have used ${\Gamma^{\alpha}_{~\mu \nu}}$ 
	in ${{\nabla}_{\mu}}$. This also agrees with a local theory based on the principle of equivalence considered below. 
	Same discussions remain valid with ordinary matter field actions: 
	$S_{M} = \int{d^4 x}{{\mathcal{L}_M}}$, where ${\sqrt{-g}}$ is included in ${{\mathcal{L}_M}}$. While deriving the corresponding Euler-Lagrange's equations, we have to eliminate the four-divergence term: ${{\nabla'}_{\mu}[{{\partial {\mathcal{L'}_M} \over {\partial ({{{\nabla'}_{\mu}}{\chi_a}})}}{\delta {\chi_a}}}]}$ by applying the Stokes' theorem. Here, ${\mathcal{L'}_M}$ is the matter Lagrangian density evaluated using the total connection ${\Theta^{\alpha}_{~\mu \nu}}$. This is not possible until ${{\partial {\mathcal{L'}_M} \over {\partial ({{{\nabla'}_{\mu}}{\chi_a}})}}{\delta {\chi_a}}}$ is a relative contravariant vector field of weight $+1$ [27,29]. Otherwise, the functional derivative $\frac{\delta S_{M}}{\delta {\chi_a}}$ can contain a non-covariant term. When we can eliminate the total divergence term, the functional derivative is given by:
	$\frac{\delta S_M}{\delta \chi_a} = - {{\nabla'}_{\mu}({{\partial {\mathcal{L'}_M} \over {\partial ({{{\nabla'}_{\mu}}{\chi_a}})}}}) + {\partial {\mathcal{L'}_M} \over {\partial {\chi_a}}}}$.

	We now discuss the possible coupling between the connection scalars $\omega,\psi$ and matter fields that can arise through the equations of motion obtained by using a variational principle. For scalar fields (including many non-linear models, [37]), ${C^{\alpha}_{~\mu \nu}}$ does not appear in the corresponding variational derivatives $\frac{\delta S_{M}}{\delta {\chi_a}}$ and field equations even if we use ${\Theta^{\alpha}_{~\mu \nu}}$ to evaluate ${{\nabla'}_{\mu}({{\partial {\mathcal{L'}_M} \over {\partial ({{{\nabla'}_{\mu}}{\chi_a}})}}})}$. Thus, we can use ${\Gamma^{\alpha}_{~\mu \nu}}$ to construct $S_M$, $\frac{\delta S_{M}}{\delta {\chi_a}}$ and field equations. A similar attribute remains valid for the gauge invariant electromagnetic field tensor described by the Lagrangian density: $- \frac{1}{4} {\sqrt{-g}} F_{\mu \nu} F^{\mu \nu}$. We get the following set of equations for free electromagnetic theory:

	\ba
	{\nabla_\mu}{F^{\mu \nu}} & = & - {C^{\nu}_{~\alpha \beta}}{F^{\alpha \beta}} = 0 \\ \nonumber
	{\nabla_{[\mu}}{F_{\nu \kappa ]}} & = & 0.
	\ea

	\noindent{The right hand side of the first equation is replaced by a current conserved \textit{w.r.t} ${{\nabla}_{\mu}}$ when we have sources. The same remains valid with nonabelian gauge theories. We have, [37]:}
	
	\ba
	F_{\mu \nu a} & = & {\nabla_\mu}{A_{\nu a}} - {\nabla_\nu}{A_{\mu a}} - {\theta^{bc}_{~~a}}{A_{\mu b}}{A_{\nu c}} \\ \nonumber
	{\nabla_\mu}{F^{\mu \nu}_{~~a}} & - & {\theta^{bc}_{~~a}}{A_{\mu b}}{F^{\mu \nu}_{~~c}} = 0, ~ 
	{\theta^{bc}_{~~a}} = - {\theta^{cb}_{~~a}}
	\ea
	
	\noindent{where ${\theta^{bc}_{~~a}}$ is the structure constant of the gauge group. Note that ${C^{\nu}_{~\alpha \beta}}(\omega,\psi)$ does not appear explicitly in ${\mathcal{L'}_M}$ of all three cases mentioned so far. We now consider the Dirac fields described by the Lagrangian density [38,39,40]:}

	\be
	\mathcal{L'}  =  -{\sqrt{-g}} \{ \frac{1}{2} [\bar{\chi} {\gamma^{\mu}} {\nabla'_\mu}{\chi} - ({\nabla'_\mu}{\bar{\chi}}){\gamma^{\mu}} \chi] + m \bar{\chi} {\chi} \}, ~ {\gamma^{\mu}} = {V^{\mu}_{a}}{\gamma^a}
	\ee
	
	\noindent{where $\chi$ are the $SL(2,c)$ spinors, ${V^{\mu}_{a}}$ are the vierbeins [39,40] and ${\gamma^a}$ are the flat spacetime Dirac matrices. Note that the vierbeins locally satisfy: ${V^{\mu}_{a}}(x) {V_{b \mu}}(x) = {\eta_{ab}}$, but ${V^{\mu}_{a}} (x)$ at two different points can not be related by parallel transport with non-trivial ${C^{\alpha}_{~\mu \nu}}$. The covariant derivative on the $SL(2,c)$ spinors $\chi$ are given by [39,40]:}

	\ba	
	{\nabla'_\mu} & = & \partial_{\mu} + {\dfrac{1}{2}}{\sigma^{bc}}{V^{\nu}_{b}}({{\nabla'}_{\mu}}{V_{c \nu}}) \\ \nonumber 
	~  & = & {\nabla_\mu} - \frac{1}{2} {\sigma^{bc}}{V_{b}^{\nu}}{V_{c \alpha}} {C^{\alpha}_{~\mu \nu}}, ~~ 
	{\sigma^{bc}} = - \frac{i}{4} [\gamma^b , \gamma^c ] \\ \nonumber
	& = & {\nabla_\mu} + C_{\mu}.
	\ea

	\noindent{For ${C^{\alpha}_{~\mu \nu}}$ given by Eq.(2), $\mathcal{L'}$ simplifies to:}
	
	\be
	\mathcal{L'}  =  -{\sqrt{-g}} \{ \frac{1}{2} [\bar{\chi} {\gamma^{\mu}}{\nabla_\mu}{\chi} - ({\nabla_\mu}{\bar{\chi}}){\gamma^{\mu}} \chi] + m \bar{\chi} {\chi}  - 
	\frac{3i}{4} (\bar{\chi} {\gamma^{\mu}} \chi) {\nabla_\mu}(\omega - \psi) \}.
	\ee
	
	\noindent{We can neglect the second term in a theory where the Dirac spinors are coupled with gauge fields through  conserved currents proportional to $\bar{\chi} {\gamma^{\mu}} \chi$. This includes quantum electrodynamics. Gauge fields are described as before. Thus, in all four cases mentioned above, we can replace ${\Theta^{\alpha}_{~\mu \nu}}$ by ${\Gamma^{\alpha}_{~\mu \nu}}$ in ${\mathcal{L'}_M}$ and express the latter as ${\mathcal{L}_M}$. We can use ${\mathcal{L}_M}$ to obtain corresponding field equations and construct matter field stress-tensors in ${\nabla_\mu}$. Such stress-tensors lead to Einstein's equation consistent with the Bianchi identity in ${\Gamma^{\alpha}_{~\mu \nu}}$ when we add them in the \textit{r.h.s} of Eq.(6).}

	Things become more complicated with weak interaction that include relative (axial) vector currents [27,37]. To illustrate, we consider the electroweak theory [37] without the complex scalar fields and use the Levi-Civita connection:
	
	\ba
	\mathcal{\tilde L} & = & {\sqrt{-g}} [- \frac{1}{4} {\vec{A}_{\mu \nu}}.{\vec{A}^{\mu \nu}} - \frac{1}{4} {{B}_{\mu \nu}}{{B}^{\mu \nu}} \\ \nonumber 
	& - & {\bar{e}_{R}}({\gamma^{\mu}}{\nabla_{\mu}} + ig'{\gamma^{\mu}}{B_{\mu}}){e_R} - {\bar{L}_e}({\gamma^{\mu}}{\nabla_{\mu}} + i \frac{g'}{2} {\gamma^{\mu}}{B_{\mu}} - i \frac{g}{2}{\sigma_i}{\gamma^{\mu}}{A^{i}_{\mu}}){L_e}] \\ \nonumber
	{L_e} & = & {\begin{pmatrix}
			\nu_L \\
			e_{L} 
	\end{pmatrix}}, ~ e_{R/L} = \frac{1}{2} (1 \pm \gamma_5){\chi_e}, ~ \nu_{L} = \frac{1}{2} (1 - \gamma_5){\chi_\nu}
	\ea
	
	\noindent{where $\chi_e$ and $\chi_{\nu}$ are the Dirac fields for the electrons and neutrinos respectively. Other symbols have usual meanings [37]. As before, ${\nabla_{\mu}}$ is evaluated using the Levi-Civita connection. We have discussed the free gauge field terms before. Here, we first consider the Dirac field terms that are only coupled with gravity. This part can be expressed as:}
	
	\ba
	{\mathcal{\tilde L}}_D & = & {\sqrt{-g}} [- {\bar{\chi}_e} {\gamma^{\mu}}{\nabla_\mu}{\chi_e} \\ \nonumber
	& ~ & - \frac{1}{2} {\bar{\chi}_\nu} {\gamma^{\mu}}{\nabla_\mu}{\chi_\nu}  
	+ \frac{1}{2} {\bar{\chi}_\nu} {\gamma^{\mu}}{\gamma_5}{\nabla_\mu}{\chi_\nu}].
	\ea

	\noindent{We do not have any problem with the electron field part. However, the variational derivative: $\frac{\delta {\tilde S}_D}{\delta {\chi_\nu}(x)}$ is non-covariant. This is because ${\sqrt{-g}}[{\bar{\chi}_\nu} {\gamma^{\mu}}{\gamma_5}({\delta {\chi_\nu}})]$ is a vector and, as mentioned before, we cannot eliminate the total divergence in $\int {\bf e} {\nabla_{\mu}}[{\sqrt{-g}}({\bar{\chi}_\nu} {\gamma^{\mu}}{\gamma_5}{\delta {\chi_\nu}})]$ completely by using the Stokes' theorem [27,29]. We can eliminate total divergence terms by using the Stokes' theorem for relative vector fields of weight $+1$. We find that: $\frac{\delta {\tilde S}_D}{\delta {\bar{\chi}_\nu}(x)} = - {\sqrt{-g}}{\gamma^{\mu}}{\nabla_\mu}{\chi_{\nu L}}(x)$ and $\frac{\delta {\tilde S}_D}{\delta {\chi_\nu}(x)} = {\sqrt{-g}}[{\bar{\chi}_{\nu L}}(x){\gamma^{\mu}}{\overleftarrow{\nabla}_\mu} + \frac{1}{2} {\Gamma^{\mu}_{~\mu \alpha}}({\bar{\chi}_\nu}{\gamma^{\alpha}}{\gamma_5})]$. This is expected because of the mixing of scalar density [26] and scalar in ${\mathcal{\tilde L}}_D$ for ${\chi_\nu}$. We also find that we cannot derive a consistent pair of equations for ${{\chi}_\nu}$ and ${\bar{\chi}_\nu}$ by applying the principle of least action directly to the above action when ${\bar{\chi}_\nu}$ is taken to be the adjoint of ${{\chi}_\nu}$ given by ${\chi_\nu^\dagger}{\gamma^0}$. The latter choice is required to construct various physical variables like stress-tensor from ${\mathcal{\tilde L}}_D$ and will be used in the present article. We do not have problems when: ${\Gamma^{\mu}_{~\mu \nu}} = 0$, which is not valid in general. We solve this problem by adding the kinetic term for $\nu_{R}$. Thus, we modify ${\mathcal{\tilde L}}_D$ to:}

	\be
	{\mathcal{L}}_D = - \frac{1}{2} {\sqrt{-g}}[{\bar{\chi}_e} {\gamma^{\mu}}{\nabla_\mu}{\chi_e} - ({\nabla_\mu}{\bar{\chi}_e}){\gamma^{\mu}}{\chi_e}
	+ {\bar{\chi}_\nu} {\gamma^{\mu}}{\nabla_\mu}{\chi_\nu} - ({\nabla_\mu}{\bar{\chi}_\nu}){\gamma^{\mu}}{\chi_\nu}].
	\ee

	\noindent{The equations for $\nu_{L}$ and ${\bar{\nu}}_L$ can be obtained from that of ${\chi_\nu}$ and ${\bar{\chi}_\nu}$. There are other important advantages. We will find below that ${\mathcal{L}}_D$ can be used to construct an expression for conserved vector current for the neutrino while this is not valid for ${\mathcal{\tilde L}}_D$. These comments remain valid in flat spacetime with curvilinear coordinates. In flat spacetime, the right-handed neutrinos can be useful to remove triangle anomalies in the lepton sector [41], and can be important in explaining neutrino oscillation [42]. They can also be relevant in the dark matter research. We now consider possible conserved currents associated with ${\mathcal{\tilde L}}_D$, ${\mathcal{L}}_D$ and the theory that we obtain when we replace ${\mathcal{\tilde L}}_D$ by ${\mathcal{L}}_D$ in Eq.(12).}

	\noindent{(\textbf{i}) We find that the global phase transformation: ${\chi} \rightarrow {e^{- i c}} {\chi}$, where $c$ is a constant, is an exact symmetry of both ${\mathcal{\tilde L}}_D$ and ${\mathcal{L}}_D$. We generate a conserved vector current for the Dirac spinors by generalizing to local phase transformation: ${\chi_{e/ \nu}} \rightarrow {e^{- i \alpha(x)}} {\chi_{e/ \nu}}$ and define the current as: ${\sqrt{-g}} {\nabla_{\mu}j^{\mu}} =  \frac{\delta S_M}{\delta \alpha(x)}$, [38]. We have: ${\nabla_{\mu}j^{\mu}} = 0$, for the classical fields. Current conservation also follows from the equations of motion directly. In an anomaly-free gauge theory minimally coupled with the Dirac fields, the current remains conserved and the action remains invariant under local phase transformation even for the non-classical configurations relevant to the quantum theory. Thus, we can define the conserved current using the above expression with $\frac{\delta S_M}{\delta \alpha(x)} = 0$, [43]. We do not have any problem with electrons. However, with ${\mathcal{\tilde L}}_D$, we cannot define a covariant functional derivative $\frac{\delta S_M}{\delta \alpha(x)}$ for the neutrinos. This is because, as before, we cannot apply the Stokes' theorem to the vector: ${\sqrt{-g}}{\alpha (x)}(\bar{\chi_\nu} {\gamma^{\mu}}{\gamma_5} \chi_\nu)$, where $\alpha(x)$ is any regular scalar field and hence, cannot define a covariant functional derivative $\frac{\delta {~~~}}{\delta \alpha(x)}$ of the functional: $\int {\bf e} {\sqrt{-g}} [{\bar{\chi}_\nu} {\gamma^{\mu}}{(1 - \gamma_5)} \chi_\nu] {\nabla_{\mu} \alpha}$. Similar comments can remain valid for curvilinear coordinates in flat spacetime. We do not have such problems with ${\mathcal{L}}_D$. The right-handed neutrinos have been used before to cancel anomalies [41]. Thus, ${\mathcal{L}}_D$ is also more appropriate to define the conserved vector current for the neutrinos.}

	\noindent{(\textbf{ii}) ${\chi} \rightarrow {e^{- i c \gamma_5}} {\chi}$ is another symmetry transformation [43] of both ${\mathcal{\tilde L}}_D$ and ${\mathcal{L}}_D$, although we cannot define axial vector currents using $\frac{\delta S_M}{\delta \alpha(x)}$ when we generalize $c$ to $\alpha(x)$. This remains valid for both ${\mathcal{\tilde L}}_D$ and ${\mathcal{L}}_D$ and is due to the reasons discussed before regarding the Stokes' theorem in curved spacetime. Axial vector currents like $\bar{\chi} {\gamma^{\mu}}{\gamma_5} \chi$ remain conserved for the classical solutions by the equations of motion obtained from ${\mathcal{L}}_D$. Various axial vector currents may remain approximately conserved in the quantum theory in the presence of gravity. However, the comments regarding Stokes' theorem in curved spacetime remain valid for curvilinear coordinates in flat spacetime. Thus, we expect anomalies for various axial vector currents in a naive interacting theory even in flat spacetime. Anomalies in the axial vector currents are common in flat spacetime when the fermions are coupled with various gauge fields through the axial vector currents [37,43]. This includes pion decay. We thus replace Eq.(12) by the following expression:}

	\ba
	\mathcal{L} & = & {\sqrt{-g}} \{ - \frac{1}{4} {\vec{A}_{\mu \nu}}.{\vec{A}^{\mu \nu}} - \frac{1}{4} {{B}_{\mu \nu}}{{B}^{\mu \nu}} \\ \nonumber 
	& - & \frac{1}{2} [{\bar{\chi}_e} {\gamma^{\mu}}{\nabla_\mu}{\chi_e} - ({\nabla_\mu}{\bar{\chi}_e}){\gamma^{\mu}}{\chi_e}
	+ {\bar{\chi}_\nu} {\gamma^{\mu}}{\nabla_\mu}{\chi_\nu} - ({\nabla_\mu}{\bar{\chi}_\nu}){\gamma^{\mu}}{\chi_\nu}] \\ \nonumber
	& + & ig'{\bar{e}_{R}}{\gamma^{\mu}}{B_{\mu}}{e_R} + i\frac{g'}{2} {\bar{L}_e}{\gamma^{\mu}}{B_{\mu}}{L_e} - i\frac{g}{2} {\bar{L}_e}{\sigma_i}{\gamma^{\mu}}{A^{i}_{\mu}}{L_e} \}.
	\ea

	\noindent{(\textbf{iii}) ${\mathcal{\tilde L}}$ is invariant under simultaneous gauge and local phase transformations of various Dirac fields with $e_L$ and $e_R$ transforming differently due to parity violating effects. The same remains valid for $\mathcal{L}$ with $\nu_R$ remaining unchanged under such transformations. However, $S_{M}$ obtained from ${\mathcal{\tilde L}}$ is not invariant under pure gauge transformations: $A_{\mu} \rightarrow A_{\mu} + {\nabla_\mu}{\lambda(x)}$, without any change in the phases of various Dirac fields. This is consistent with (i) and (ii) and indicates  possible axial current anomaly. The same remains valid for $\mathcal{L}$. Note that for a pure gauge transformation without any change in the phases of Dirac fields, the variation of the action comes from the coupling terms between the potentials and fermionic bilinear forms: $\bar{\chi} {\gamma^{\mu}}M \chi, ~ M = 1, {\gamma_5}$. As before, Stokes' theorem implies that $\frac{\delta {S_M}}{\delta \lambda(x)}$ is not tensorial in general under pure gauge transformations even if we impose $\nabla_{\mu}(\bar{\chi} {\gamma^{\mu}} \chi) = 0$. Thus, with $\mathcal{L}$ we cannot define conserved currents by using: $\frac{\delta S_M}{\delta \lambda(x)}$. This is the reason that we have started with invariance under phase transformations of Dirac fields to define possible conserved currents [43] in the present theory rather than the gauge transformations. A similar aspect remains valid with curvilinear coordinates in flat spacetime and we can have axial vector current anomaly in flat spacetime as mentioned before. The above discussions indicate that various regularization schemes used in quantum theory need to preserve pure gauge transformation and corresponding local phase transformations in the Dirac fields mentioned before. We can still have anomalies in axial vector currents that depend on $\mathcal{L}$, regularization schemes and possibly boundary conditions [43,44,45]. (\textbf{i}), (\textbf{ii}) and (\textbf{iii}) can be important in the early universe.}

	We now replace the Levi-Civita connection in Eq.(15) by ${\Theta^{\alpha}_{~\mu \nu}}$. As in the case of Eq.(9), it is easy to show that $\omega$ and $\psi$ can be eliminated from the action obtained from $\mathcal{L}$.
	This helps us to take $\mathcal{L'}({\Theta^{\alpha}_{~\mu \nu}}) = \mathcal{L}({\Gamma^{\alpha}_{~\mu \nu}})$ for the Dirac fields in the electroweak theory with $\mathcal{L}$ suitably generalized from Eq.(15) by adding the complex scalar fields [37]. Thus, we can construct the standard model using ${\nabla_\mu}$ only. ${C^{\alpha}_{~\mu \nu}}(\omega,\psi)$ given by Eq.(2) does not appear explicitly in action for scalar fields, abelian and nonabelian gauge fields. It also does not appear in the equations of motion for these fields even if we include it in the corresponding covariant derivatives. For spin half spinors, ${C^{\alpha}_{~\mu \nu}}(\omega,\psi)$ only contributes boundary terms when we use Eq.(14).
	${C^{\alpha}_{~\mu \nu}}(\omega,\psi)$ will not be present in matter field stress-tensors also. This partly justifies our efforts to explain dark energy using $\psi$. We conclude that we do not have any problem with the contracted Bianchi identity when we generalize the standard model to curved spacetime described by the connection ${\Theta^{\alpha}_{~\mu \nu}}$ provided we modify the action so that we do not have any problem with ${\Gamma^{\alpha}_{~\mu \nu}}$. The right-handed neutrino, $\nu_{R}$, remains a free field and can be important for various purposes mentioned before, including QFT in flat spacetime. We can follow (\textbf{3}) below when we use Eq.(12) to describe electroweak theory. Otherwise, we have coupling between $\omega$, $\psi$ and neutrinos that can give novel features in the semiclassical theory, quantum fields in curved spacetime. In this case, we have to consider a particular combination of $\omega$ and $\psi$ to have a stable theory. This is discussed in the next section. We can not define conserved vector current for the neutrinos even when $\omega, \psi = 0$. We can have coupling between $(\omega, \psi)$ and general tensor fields that are not antisymmetric. In this case, we can again follow (\textbf{3}) below. It is unexpected that $\omega$ and $\psi$ will couple with classical sources if they do not couple with quantum fields. The stress tensor of such sources, when added to the source of Einstein's equation given by Eq.(6), gives equation of motion consistent with the Bianchi identity [29]. Torsion will be discussed later.

	{~}

	(\textbf{3}) \textbf{Restricting ${{\Theta}^{\alpha}_{~\mu \nu}}$ to ${\Gamma^{\alpha}_{~\mu \nu}}$ in matter and radiation actions}: 
	
	According to the principle of general covariance, we can restrict ${{\Theta}^{\alpha}_{~\mu \nu}}$ to ${\Gamma^{\alpha}_{~\mu \nu}}$, when we need to replace the partial derivatives acting on matter and radiation fields
	by covariant derivatives. When we try to construct a theory of gravity with sources, of the form given by Eq.(6), we supplement the \textit{r.h.s} of Eq.(6) with stress tensors that are symmetric and conserved \textit{w.r.t} ${\nabla_\mu}$. The stress tensor of ideal fluid does not contain the connection and is conserved \textit{w.r.t} ${\Gamma^{\alpha}_{~\mu \nu}}$. Quantum fields are introduced in flat spacetime as different irreducible representations of $SL(2,c)$, parity and various internal symmetries. Corresponding actions are generalized to local Lorentz invariance in curved spacetime to couple with gravity. This is according to the principle of equivalence. Gauge theories are described by gauge covariant derivatives of potentials that do not contain the connection [31]. Partial derivatives acting on $SL(2,c)$ spinors are generalized from flat spacetime to curved spacetime by replacing them with suitable combinations of covariant derivatives given by ${\nabla'_{\mu}}$, ${V^{\mu}_{a}}{\partial_{\mu}}$, ${V^{\mu}_{a}}{\nabla'_{\mu}}$ and ${{\nabla'}_{\nu}}{V^{\mu}_{a}}$ where ${V^{\mu}_{a}}$ are the vierbeins. The choice of connection depends on the contracted Bianchi identity. In general, stress tensor for $SL(2,c)$ spinors are always conserved \textit{w.r.t} ${{\nabla'}_{\mu}}$ when the connection in ${{\nabla'}_{\mu}}$ is ${\Gamma^{\alpha}_{~\mu \nu}}$, [40]. ${\Gamma^{\alpha}_{~\mu \nu}}$ is also used in the Newman-Penrose formalism. Thus, a theory where we do not include ${C^{\alpha}_{~\mu \nu}}(\omega,\psi)$ in the covariant derivatives when acting on matter and radiation from the beginning, is a possible theory.

	{~}

 We find from the above three points that it is nontrivial to extend the standard model to curved spacetime even with the Levi-Civita connection. We have to introduce the right-handed neutrino as a free field singlet in the standard model to have a theory with conserved vector current for the neutrinos. Observations in (\textbf{2}) and (\textbf{3}) imply that for fields relevant to the standard model, fields that are various representations of $SL(2,c)$ with the addition of the right handed neutrinos, we can use only ${\Gamma^{\alpha}_{~\mu \nu}}$ as the connection present in the covariant derivatives acting on them even when $C^{\alpha}_{~\mu \nu}(\omega,\psi)$ is finite. Variables like stress-tensors of ordinary matter and radiation fields also contain only ${\Gamma^{\alpha}_{~\mu \nu}}$ as the connection. We find that this agrees with: (i) a principle of least action, (ii) principle of general covariance, (iii) Stokes' theorem, (iv) contracted Bianchi identity in presence of finite non-metricity and (v) principle of equivalence.  We do the same to defining the gauge invariant electromagnetic field tensor in the presence of torsion [31]. This explains why Levi-Civita connection and Einstein's gravity explain the observable universe to a very good approximation. 
Under these circumstances, $C^{\alpha}_{~\mu \nu}(\omega,\psi)$ only contributes a stress-tensor in the \textit{r.h.s}
of the modified Einstein's equation given by Eq.(6) that presently has small effect on spacetime.

Depending on $\psi$, we may not have: ${V^{\mu}_{a}}{V_{b \mu}} = {\eta_{ab}}$, where ${\eta_{ab}}$ is the flat spacetime metric, when we parallel transport ${V^{\mu}_{a}}$ from one point to a distant point in the spacetime manifold [23,24,25]. Thus, we do not have any problem with the principle of equivalence, but the global splitting of spacetime into space and time may not remain exact. In the Einstein-Palatini formalism considered here, $\omega$ and ${\psi}$ obey the homogeneous wave equation and can be finite everywhere \textit{i.e} they are non-localized. We can add suitable potential terms for themselves. We will discuss this later. Dark energy does not interact with ordinary matter directly and is observed through large scale cosmological observations. $\psi$ contributes a negative source stress tensor to Einstein's equation and the preceding discussions indicate that $\omega$ and ${\psi}$ need not be present in the covariant derivatives acting on matter and radiation fields. Thus, $\omega / \phi$ and $\psi$ need not to have ordinary matter and radiation as their sources [31]. All these make $\omega$ and $\psi$ as possible candidates to explain inflation and dark energy. This qualitatively agrees with the energy budget mentioned in section:I. It is unexpected that $10 \%$ ordinary matter can produce $70 \%$ dark energy. 
This may be required to construct a stable theory. We will discuss this issue in the next section.
We find that a part of dark matter can be the right-handed neutrinos. ${\phi}$ is similar to the dilaton [46].  Cosmological observations indicate that non-metricities produced by $\omega$ and ${\psi}$ are presently very small. This is consistent with the smallness of the cosmological constant in the $\Lambda$CDM model. $\omega$ and $\psi$ manifest themselves through long range geodesics and the global structure of spacetime. However, local inhomogeneities and anisotropies in $\omega$ and $\psi$ can cause significant effects on geodesic motions in the corresponding regions. 
$\omega$ and $\psi$ are useful to discuss geometry.
In the next section, we will find that it is more useful to consider linear combinations of $\omega$ and $\psi$ in discussing stability and their effects like inflation and dark energy.

	\section*{IV. Extension of Einstein's Theory of Gravity}

	We can use classical theories and quantum field theory in curved spaces to find the effects  
	of ${\omega}$ and ${\psi}$. This will be useful in cosmic epoch. We now compare Eqs.(4,5) with the action of scalar-tensor theories [6]:

	\be
	S = {\int {\bf e}}{\sqrt{-g}}{\big[}f(\phi, R) - \zeta(\phi){({\nabla}{\phi})^{2}}{\big]}
	+ {S_{m}(\chi_a , \nu_R , {g_{\mu \nu}})} 
	\ee

	\noindent{where $\chi_a$ is the collection of ordinary matter and radiation. Eq.(5) indicates that: $\zeta(\omega) = {3 \over 2}, ~~ \zeta(\psi) = - 3$. $f(\phi, R) = R$ for both fields. Thus, the scalars mentioned at the beginning of the first section like quintessence and k-essence can be related to quantum gravity and can have geometrical origin. When required, we can modify the theory by adding suitable $\omega$ and $\psi$ dependent terms in Eq.(5). We get:}

	\be
	\mathcal{L} = {\sqrt{-g}} (\mathcal{R'} - \mathcal{V})
	\ee

	\noindent{where $\mathcal{V}$ contains the added terms which should be consistent with: ${{\nabla}^{\alpha}}{{G}_{\mu \alpha}} = 0,$ if we express the equations in the form similar to Eq.(6). In general, the above Lagrangian density will give coupled equations for $\omega$ and $\psi$. We obtain a decoupled theory by diagonalizing the symmetric bilinear form in ${{\nabla}^{\mu} {\omega}}$ and ${{\nabla}^{\nu} {\psi}}$ given by Eq.(5), [47]. This bilinear form is similar to: ${a^2} - 2ab - 2{b^2}$ and the corresponding symmetric matrix has eigenvalues: ${\lambda_{\pm}} = (-1 \pm \sqrt{13})/2$ . $\mathcal{R'}$ is now given by the following expression:
		
		\ba
		{\mathcal{R'}} & = & {R}  - {3 \over 2}[{({\nabla}{\omega})^{2}} 
		- 2{({\nabla}{\psi})^{2}} - 2 ({{\nabla}_{\kappa}{\omega}})({{\nabla}^{\kappa}{\psi}})] \\ \nonumber
		& = & {R}  - {3 \over 2}{g^{\mu \nu}}[{\lambda_{+}}{({{\nabla}_{\mu}{\Phi'}})({{\nabla}_{\nu}{\Phi'}})}
		+ {\lambda_{-}}{({{\nabla}_{\mu}{\Psi'}})({{\nabla}_{\nu}{\Psi'}})}] 
		\ea

		\noindent{where $\Phi'$ and $\Psi'$ are related to $\omega$ and $\psi$ by the following expression:}

		\be
		{\begin{pmatrix}
				\Phi' \\
				\Psi' 
		\end{pmatrix}}  = 
		{S^T} {\begin{pmatrix}
				\omega \\
				\psi 
		\end{pmatrix}}, ~~~ S = {\begin{pmatrix}
				{c^{+}_{1}} & {c^{-}_{1}} \\
				{c^{+}_{2}} & {c^{-}_{2}} 
		\end{pmatrix}}.
		\ee

		\noindent{Here ${c^{\pm}_{i}}$ are the components of two orthonormal eigenvectors. $S$ is given by the following expression:}

		\be
		S = 
		{\begin{pmatrix}
				{\sqrt{4 \over {26 - 6\sqrt{13}}}} &  {\sqrt{4 \over {26 + 6\sqrt{13}}}} \\
				{{3 - \sqrt{13} \over 2}}\sqrt{4 \over {26 - 6\sqrt{13}}} & {{3 + \sqrt{13} \over 2}}\sqrt{4 \over {26 + 6\sqrt{13}}} 
		\end{pmatrix}}.
		\ee

		\noindent{We can absorb the positive
			numerical factors into $\Phi' ,\Psi'$ and the modified curvature scalar is given by:}
		
		\be
		{\mathcal{R'}} = {R}  - {1 \over 2}[{({{\nabla}{\Phi}})^2}
		- {({{\nabla}{\Psi}})^2}].
		\ee

		\noindent{The above expression gives a more convenient way to discuss the physical effects of the non-metricity scalar fields $\omega$ and $\psi$. We can express $C^{\alpha}_{~\mu \nu}$ in terms of $\Phi$ and $\Psi$ using the inverse of Eq.(19). We find that $\Psi$ contributes a negative stress tensor to Einstein's equation and is a phantom scalar [6] of geometric origin. We always get a scalar field with a negative stress tensor in Einstein's equation, irrespective of the initial linear combinations of $\nabla^{\mu} \omega$ and $\nabla^{\nu} \psi$, provided they give finite $C^{\alpha~\mu}_{~\mu}$. $\Psi$ vanishes when $C^{\alpha}_{~\mu \nu}$ is traceless in the lower indices. We can introduce various potential terms like ${{\mathcal V}_1}(\Phi)$, ${{\mathcal V}_2}(\Psi)$ and as per requirement. ${{\mathcal V}_1}(\Phi)$ and ${{\mathcal V}_2}(\Psi)$ can give us a theory more general than that given by Eq.(16) and new effects in the scalar-tensor theories. The contracted Bianchi identity will remain valid for this theory. Thus, transforming back to $(\omega, \psi)$ we will get a theory with $(\omega, \psi)$ that preserves the contracted Bianchi identity.}

		We can consider a theory containing $\Phi$, $\Psi$ and the cosmological constant $\Lambda$. This will lead to a possible theory of inflation together with dynamic and spatially varying dark energy. Einstein's equation will be generalized to:
		
		\be
		G_{\mu \nu} + {\Lambda}{g_{\mu \nu}} - {\Lambda_{\mu \nu} (\Phi, \Psi)} = 8{\pi}{P_{\mu \nu}(\chi_a )} 
		\ee

		\noindent{where ${\Lambda_{\mu \nu} (\Phi, \Psi)}$ includes ${P_{\mu \nu}(\Phi)}, {P_{\mu \nu}(\Psi)}$ and potential terms for $\Phi$ and $\Psi$. The left hand side gives a generalization of the $\Lambda$CDM theory linear in curvature and consistent with the Bianchi identities. This is along the line of introducing the cosmological constant in Einstein's equation. ${{P}_{\mu \nu}(\chi_a)}$ is the stress tensor of ordinary matter and radiation including the right handed neutrinos.  We can construct new non-trivial solutions of the vacuum Einstein's equation [29,48,49], \textit{i.e}, for ${P_{\mu \nu}(\chi_a )} = 0$, with different choices of ${\Lambda_{\mu \nu} (\Phi, \Psi)}$.}

		An important issue is the stability of the theory. $\Psi$ provides a negative kinetic term, thus behaving like a phantom scalar [5,50,51,52,53]. Phantom field is a possible candidate to explain dark energy [5,6]. 
		A phantom field, interacting with ordinary matter and radiation, can lead to instability both in the classical and quantum theory. Stable classical theories are discussed by a few authors by adding suitable potential terms for the phantom fields [52,53]. There remains the issue of unstable vacuum [54,55]. In the present theory, $\Psi$ does not couple with ordinary matter and scattering between the two sectors causing instability of the vacuum is not possible. Processes of the form: $Vacuum \rightarrow 2 \tilde{\gamma} + 2 \tilde{\psi}$, where $\tilde{\gamma}$ is a photon and $\tilde{\psi}$ gives a phantom quanta need careful analysis. Here, we illustrate this with the simpler case of the process: $Vacuum \rightarrow 2 \mathcal{\tilde{\phi}} + 2 \tilde{\psi}$, where $\tilde{\phi}$ is a $\Phi$ field quantum. Firstly, we don't allow negative energy states for the phantom quantum $\tilde{\psi}$. Such states don't exist in a quantum field theory. Also, such states won't contribute negatively to the total energy in the matter field part of Eq.(22) and the above process is not possible. These processes use the kinetic terms of the form: ${({{\nabla}{\Psi}})^2}$ and a graviton propagator [54]. In a perturbative theory of spin-two particle in flat spacetime background, the non-vanishing matrix elements are of the form
		$\langle \tilde{\phi} \tilde{\phi} \tilde{\psi} \tilde{\psi}|(a_{\tilde{\phi}} a_{\tilde{\phi}} a_{\tilde{\psi}} a_{\tilde{\psi}})^{\dagger}|0\rangle$ and are obtained from:
		$\left\langle \tilde{\phi} \tilde{\phi} \tilde{\psi} \tilde{\psi}|\int \int {d^4 x} 
		{d^4 y}:{{\nabla}^{\alpha}{\Psi}}{{\nabla}^{\nu}{\Psi}}(x)S^{F}_{\alpha\beta,\mu\nu}(x-y)
		{{\nabla}^{\mu}{\Phi}}{{\nabla}^{\nu}{\Phi}}(y):|0\right\rangle$, where $S^{F}_{\alpha\beta,\mu\nu}(x-y)$ gives the translationally invariant Feynman propagator for the spin -two graviton in flat spacetime. This requires negative energy states for both $\tilde{\phi}$ and $\tilde{\psi}$. Thus, the decay: $Vacuum \rightarrow 2 \tilde{\phi} + 2 \tilde{\psi}$ is not possible in a perturbative quantum field theory. Replacing $\Phi$ by higher spin fields won't change the situation. Similar situation will remain valid in any stationary background. In addition, such processes break conservation of linear momentum in a spacetime translationally invariant in one or more spatial direction(s). This is the case with the flat spacetime. In a non-perturbative theory, the coupling between $g_{\mu \nu}$ and $\Psi$, given by ${({{\nabla}{\Psi}})^2}$, can be put in the form ${\Psi}{\nabla^2}{\Psi}$, vanishes for the classical solutions, thus prohibiting the decay of vacuum to a detectable on-shell spectrum. Total derivatives in the action have no effect on the classical states. ${\Psi}{\nabla^2}{\Psi}$ becomes a polynomial in $\Psi$ with a ${{\mathcal V}(\Psi)}$ for the on-shell spectrum in a non-perturbative theory, and there is no coupling with the graviton. Thus, classically stable self-interacting theory of $\Psi$ can give a stable quantum theory. A theory with static or nearly static $\Psi$ can be useful for discussing dark energy. A scalar field with negative energy density is also required to construct the \textit{steady state} model of the expanding universe [48,56,57,58,59].

		$\Phi$ contributes a positive stress tensor and gives a stable theory. Both ${Q_{\mu}}$ and ${{\bar{Q}}_{\mu \alpha \beta}}$ are finite for $\Phi$. Thus, we can consider a quantum theory with only $\Phi$ present as the non-metricity field. In the next section we will consider theories that can contain more than one $\Phi$ fields. Such a theory can be useful to explain both inflation and dark energy [20]. We can introduce quadratic self-interaction term in $\Phi$. We can couple $\Phi$ with ordinary matter by adding suitable interaction terms. Being of geometric origin, the coupling between $g_{\mu \nu}$ and $\Phi$ does not involve $G$. The same remains valid for $\Psi$. $\Phi$ and $\Psi$ are quantum gravitational in origin [23,24], and it is expected that they can play important roles in the very early universe. $g_{\mu \nu}, \Phi, \Psi$ should be treated similarly in discussing quantum effects like vacuum fluctuations. Various free field configurations of $\Phi$ and $\Psi$, their vacuum energies, quantum fluctuations and classical configurations with possible interactions can be useful to explain inflation and dark energy [6,35,50].

		Equation (22) leads to a generalization of the $\Lambda$CDM theory. We do not obtain the Newtonian theory in the weak field and slow motion limit when $\Lambda \neq 0$, [29]. This implies that $\Lambda$ is small. This is similar to the non-metricities produced by the pair ($\omega$, $\psi$) or ($\Psi$, $\Phi$) as discussed in the previous section. In this context, $\Phi$ and $\Psi$ have the advantage of being dynamical and hence can produce significant classical and quantum effects even in a free theory. This is along the line of perturbative quantum field theory and Casimir effect. Various non-trivial vacuum solutions of Einstein's equation, with and without $\Lambda$ [48,49], imply that affine connection need not to have ordinary matter and radiation as its sources even in the classical theory. The same can remain valid for $\Phi$ and $\Psi$. This agrees with the comments in the previous paragraph and the observed energy composition of our universe if these fields are useful to explain inflation and dark energy. It is unexpected that $10 \%$ ordinary matter can produce $70 \%$ dark energy. We find that inflation and dark energy possibly indicate that we have to generalize the classical structures of spacetime in the quantum domain. This is not completely unexpected. A complete theory of quantum mechanics can be partly spacetime independent algebraic theory as suggested by Einstein and apparent in various quantum entanglement experiments [60,61,62]. We will discuss applications of Eq.(22) in cosmology in a forthcoming article. ${P_{\mu \nu}(\chi_a)}$ will be multiplied by appropriate factors when we transform to non-geometrized ordinary units. Quantum corrections to the effective actions of various theories mentioned above, like that given by Eq.(22), will contain higher order curvature invariants evaluated using the Levi-Civita connection together with quantum corrections coming for various matter fields and possibly for non-metricity fields $\Phi$ and $\Psi$. The last set will depend on the requirement of potential terms for $\Phi$ and $\Psi$. This is the case with some low energy effective actions of string theory that contain ghost field [6]. Lastly, a possible approach to explaining dark energy could be to consider $\Psi$ as the possible source of $\Lambda$. Discussions on possible sources of $\Lambda$ for constant curvature spaces can be found in [48]. These are similar to $\Psi$. We cannot generate the cosmological constant with perfect fluids for which both energy density and pressure are positive definite. We will discuss this aspect later.

\section*{V. Additional Fields}

In Sect.II, we found that we can introduce scalar fields through 
${C^{\alpha}_{~\mu \nu}}$ without reference to any symmetric second rank tensors like ${\tilde{g}_{\mu \nu}}$ discussed in this article or $a_{\mu \nu}$ and $q_{\mu \nu}$ discussed previously [24,25]. It is thus possible to introduce additional scalar fields depending on observations, i.e, when such scalars couple differently with other fields including $\omega , \psi$. We can take the corresponding part of ${C^{\alpha}_{~\mu \nu}}$ to be traceless in the mixed indices or in the lower indices. Considering various models to explain inflation and dark energy discussed in [6,32,33], we can generalize Eq.(2) to the following expression:

\ba
{C^{\alpha}_{~\mu \nu}} & = & {{\delta}^{\alpha}_{~ (\mu}}{{\nabla}_{\nu)}}{\omega}
- {1 \over 2}{g_{\mu \nu}}{{{\nabla}^{\alpha}}{\omega}}\\ \nonumber
& + & {g_{\mu \nu}} {\nabla^{\alpha}}{\psi} + 
{{\delta}^{\alpha}_{~ (\mu}}{{\nabla}_{\nu)}}{\eta}
- {1 \over 4}{g_{\mu \nu}}{{{\nabla}^{\alpha}}{\eta}} \\ \nonumber
{\mathcal{R'}} & = & {R(\Gamma^{\alpha}_{~\mu \nu})}  - {1 \over 2}[{\sum_{1}^{3}} (-1)^{s_i} {(\nabla \Phi_i)}^2]
\ea

\noindent{here $s_i = 0/1$. Two of $s_i$ are zero and two are $1$, \textit{i.e}, corresponding $\Phi_i$ are phantom fields. As before, we do not need to include ${C^{\alpha}_{~\mu \nu}}({\Phi}_i)$ in the matter and radiation field actions when we include the right handed neutrinos. We can introduce self interactions depending on observations. We can also couple those $\Phi_i$ with positive kinetic terms to ordinary matter. The above model resemblances various models to address inflation and dark energy, including the multi-scalar theory of inflation [7,35,50], k-essence and phantom fields to explain dark energy, steady-state model of the expanding universe [48,56,57] and re-bouncing model of the universe [48], as special cases. The theory with only $\Phi$ and $\Psi$ associated with two traces of ${C^{\alpha}_{~\mu \nu}}$ appears to be mathematically most straightforward.}

We now consider the case of higher spin ($j > 0$) fields that can be introduced using ${C^{\alpha}_{~\mu \nu}}$
symmetric in the lower indices by expressing the latter as derivatives of lower rank tensors. In flat spacetime, second rank symmetric traceless tensors give $(A,A)$ type representations of higher spin fields like $j = 1,2$ [38]. 
We will show in the next section that we can construct a $(A,A)$ type quantum field theory for massless vector fields using symmetric traceless derivatives. Such derivatives are not gauge invariant and cannot couple with matter fields in a theory that preserves local internal gauge invariance. They can be useful for describing dark energy and dark matter. The massless higher spin particles can also be useful to construct higher spin driven inflation theories [7,8,9] in the absence of spontaneous symmetry breaking that need not to be gauge theories. We consider the case that includes a second rank symmetric traceless field ${{\bar{q}_{\mu \nu}}}$ [23,24,25]:

\be
{{C}^{\alpha}_{~\mu \nu}} = {{\delta}^{\alpha}_{~ (\mu}}{{\nabla}_{\nu)}}{\omega}
- {1 \over 2}{g_{\mu \nu}}{{{\nabla}^{\alpha}}{\omega}} + 
{g_{\mu \nu}}{{\nabla}^{\alpha} {\psi}} 
+ {\nabla^{\alpha}}{{\bar{q}_{\mu \nu}}}.
\ee

\noindent{Regarding dimensions, we have: $[{{\bar{q}_{\mu \nu}}}] = [g_{\mu \nu}]$. We have discussed in Sect.II that after we have removed the total divergence terms in $R'$ and expressed the Einstein's equation in the form given by Eq.(6), where the contribution of the non-metricity fields appears as source terms, the resulting equations have to be consistent with the contracted Bianchi identity in ${\Gamma^{\alpha}_{~\mu \nu}}$. This, together with the equations of motion, imposes severe restrictions on ${{\bar{q}_{\mu \nu}}}$. The exception being the scalar fields mentioned before. Thus, we have to modify the Einstein-Palatini action by adding suitable ${{\bar{q}_{\mu \nu}}}$ dependent terms so that the resulting theory is consistent with the contracted Bianchi identity. We can also use the metric-affine $f(R')$ theories. This may indicate that the resulting theory is described by the low energy effective action of a more fundamental theory. We can replace ${\bar{q}_{\mu \nu}}$ with the traceless derivative: $t_{\mu \nu} = {{\nabla_{(\mu}}{{B}_{\nu)}}}$ of a vector field $B_{\alpha}$ with $\nabla_{\alpha} B^{\alpha} = 0$, to have a spin one field from geometric sector. We then have:}

\be
{{C}^{\alpha}_{~\mu \nu}} = {{\delta}^{\alpha}_{~ (\mu}}{{\nabla}_{\nu)}}{\omega}
- {1 \over 2}{g_{\mu \nu}}{{{\nabla}^{\alpha}}{\omega}} + 
{g_{\mu \nu}}{{\nabla}^{\alpha} {\psi}} 
+ {\nabla^{\alpha}}{t_{\mu \nu}}
\ee

\noindent{where ${t_{\mu}^{~\mu}} = \nabla . B = 0$. Note that we have used a symmetric field $t_{\mu \nu}$ to introduce a spin one field and this field will not give a symmetric ${{R'}_{\mu \nu}}$ although the corresponding ${\mathcal{R'}_{\mu \alpha}}$ will be symmetric. We again have a problem with the contracted Bianchi identity similar to that mentioned before. We will later address this problem. We also note that we cannot eliminate $B^{\alpha}$ from the action of the Dirac field spinors and the latter can couple with $B^{\alpha}$. Such coupling is yet to be observed. Thus, the two-scalar model considered in this article is more consistent with the observations.}

We can introduce another scalar field when ${C^{\alpha}_{~\mu \nu}}$ is not symmetric in the lower indices. This can be associated with the trace of corresponding torsion tensor. We generalize Eq.(2) to the following expression:

\ba
{{\Theta}^{\alpha}_{~\mu \nu}} & = & {\Gamma^{\alpha}_{~\mu \nu}} + {{C}^{\alpha}_{~\mu \nu}} \\ \nonumber
{{C}^{\alpha}_{~\mu \nu}} & = & {{\delta}^{\alpha}_{~ (\mu}}{{\nabla}_{\nu)}}{\omega}
- {1 \over 2}{g_{\mu \nu}}{{{\nabla}^{\alpha}}{\omega}} + 
{g_{\mu \nu}}{{\nabla}^{\alpha} {\psi}} + {{\delta}^{\alpha}_{~ [\mu}}{{\nabla}_{\nu]}}{\rho}, ~~ \omega = ln(1 + \phi).
\ea

\noindent{To discuss the geometric significance of the above connection we use the non-metricity tensor ${Q_{\mu \alpha \beta}}$ [31]:}

\be
{Q_{\mu \alpha \beta}} = - {{\nabla'}_{\mu}}{g_{\alpha \beta}}
\ee

\noindent{where ${\nabla'}_{\mu}$ is evaluated using the complete connection: ${\Gamma^{\alpha}_{~\mu \nu}} + {{C}^{\alpha}_{~\mu \nu}}$. We have the following expressions for ${Q_{\mu \alpha \beta}}$:}

\be
{Q_{\mu \alpha \beta}}
= {g_{\alpha \beta}}{{\nabla}_{\mu}{\omega}} + 
2{g_{\mu (\alpha}}{{\nabla}_{\beta )}{\psi}} + {g_{\mu (\alpha}}{{\nabla}_{\beta )}{\rho}} - {g_{\alpha \beta}}{{\nabla}_{\mu}{\rho}}.
\ee

\noindent{It is convenient to express ${Q_{\mu \alpha \beta}}$ as the sum of a trace and trace-free part in the last two indices [31]:}

\be
{Q_{\mu \alpha \beta}} = {Q_{\mu}}{g_{\alpha \beta}} + {{\bar{Q}}_{\mu \alpha \beta}}.
\ee

\noindent{${{\bar{Q}}_{\mu \alpha \beta}}$ is traceless in the last two indices. ${{\bar{Q}}_{\mu \alpha \beta}}$ vanishes when $\psi, \rho = 0$. Corresponding ${{C}^{\alpha}_{~\mu \nu}}$ preserves the light cone under parallel transport due to the reparameterization invariance of the form of the geodesic equation: $t^{\mu}{{\nabla}_{\mu} t^{\nu}} = {f}t^{\nu}$, where $f$ is a scalar function on the curve [25,29]. Both trace and traceless parts of ${Q_{\mu \alpha \beta}}$ are finite for $\psi$ and $\rho$. Depending on $\psi$ and $\rho$, corresponding connection may not preserve the light cone under parallel transport [31]. Global $(3 + 1)$ -splitting of the complete spacetime manifold into space and time may not exist under these circumstances. The modified curvature scalar is now given by the following expression:}

\ba
{\mathcal{R'}} & = & {g^{\mu \alpha}}{\mathcal{R'}_{\mu \alpha}}
= {g^{\mu \alpha}} \left\lbrace  {{R}_{\mu \alpha}} 
+ 2 [{C^{\lambda}_{~[\mu |\alpha|}} {C^{\kappa}_{~\kappa |\lambda|]}}] \right\rbrace  \\ \nonumber
& = & {R} - {3 \over 2}{({\nabla}{\omega})^{2}} 
+ 3{({\nabla}{\psi})^{2}} + {3 \over 4}{({\nabla}{\rho})^{2}} 
+ 3({{\nabla}_{\kappa}{\omega}})({{\nabla}^{\kappa}{\psi}}) + 
{1 \over 2}({{\nabla}_{\kappa}{\omega}})({{\nabla}^{\kappa}{\rho}}).
\ea

\noindent{Note that ${\mathcal{R'}_{\mu \alpha}}$ is no longer symmetric. We can again diagonalize and rescale various scalar fields to obtain:}

\be
{\mathcal{R'}} = {R}  - {1 \over 2}[{({{\nabla}{\Phi}})^2}
- {({{\nabla}{\Psi}})^2} - {({{\nabla}P})^2}]
\ee

\noindent{We find that $\rho$ contributes another phantom field. $\Phi, \Psi$ and $P$ are given by linear combinations with different strengths of all three fields: $\omega$, $\psi$ and $\rho$. $\rho$ does not couple with scalar fields and Dirac fields when the Lagrangian of the latter is given by expressions similar to Eq.(14). However, $\rho$ couples with the gauge fields. This is evident from Eq.(7) and is expected in presence of torsion [31]. The field tensor is given by: 
	${F^{\alpha \beta}} = {g^{\alpha \mu}}{g^{\beta \nu}}{F_{\mu \nu}}
	= {g^{\alpha \mu}}{g^{\beta \nu}}
	[{{\nabla_{\mu}}{A_{\nu}}} -  {{\nabla_{\nu}}{A_{\mu}}} 
	+ {T^{\alpha}_{~\mu \nu}}{A_{\alpha}}]
	= {g^{\alpha \mu}}{g^{\beta \nu}}[{{\partial}_{\mu}}A_{\nu} -  
	{{\partial}_{\nu}}{A_{\mu}}]$.
	Here: ${T^{\alpha}_{~\mu \nu}} = 2{{\delta}^{\alpha}_{~ [\mu}}{{\nabla}_{\nu]}}{\rho}$ and $- {C^{\nu}_{~\alpha \beta}}{F^{\alpha \beta}} = - {F^{\nu \beta}}{\nabla_{\beta}}{(\rho)}$. The coupling of ${F^{\alpha \beta}}$ with torsion violates the current conservation in an interacting theory. There is no problem in a theory with free gauge fields. Vacuum effects can again be useful to explain inflation, dark energy and related problems [6,32,33,63,64,65]. Higher spin fields can also be introduced using torsion potentials [66].}

\section*{VI. Higher Spin Massless Bosons and Little Group}

Here, we consider the issue of constructing a semi-classical limit of the theories considered in this article. One of which can be a locally Lorentz invariant quantum field theory in curved spaces including $t_{\mu \nu}$ discussed in Sect.III. We discuss this with reference to little groups in Minkowski space and $(A,A)$ type fields [38]. We do not have any problem for the massless scalar fields for which $A = 0$, [38]. Massless integer spin particles are usually described by $(A,0) \oplus (0,A)$ type fields [38,67,68]. Here, $A$ is an integer. The most general candidate to give the little group (stabilizer) for massless particles with spin greater than zero in Minkowski space is taken to be: $W = ISO(2)$, [38]. This forces us to use $(1,0) \oplus (0,1)$ type gauge invariant antisymmetric derivatives to describe massless spin-one particles [38]. A similar situation remains valid with other massless integral spin particles. However, $ISO(2)$ is not represented by normal matrices [69] when acting on four vectors/tensors. In the following, we will find that we have to use a one parameter subgroup of $SO(3)$ as the little group for massless higher spin fields in Minkowski space. This allows us to construct a $(1,1)$ type quantum field theory for massless vector fields using symmetric traceless derivatives in addition to $(1,0) \oplus (0,1)$ type antisymmetric derivatives. Thus, we can use the symmetric traceless derivative: ${t_{\mu \nu}} = {\partial_{(\mu} B_{\nu)}}$ with ${\partial}.B = 0$ to introduce a massless spin-one particle.

In flat spacetime, the standard momentum $k^{\mu}$ used to construct the little group for massless fields is taken to be $(0,0,k,k)$, where the fourth component represents time and the metric is of signature $+2$, [38]. The little group that keeps the standard momentum unchanged is given by $ISO(2)$ and for spacetime tensors, it is taken to be $W(\theta, \alpha, \beta) = S(\alpha, \beta) R_{z}(\theta)$, [38]. Where $R_{z}(\theta)$ is a spatial rotation about the Z -axis by angle $\theta$ and $S(\alpha, \beta)$ is given by:

\be
{S^{\mu}_{~\nu}(\alpha,\beta)} = 
\begin{pmatrix}
	1 & 0 & -\alpha & \alpha\\
	0 & 1 & -\beta & \beta \\
	\alpha & \beta & 1 - \gamma & \gamma \\
	\alpha & \beta & - \gamma & 1 + \gamma
\end{pmatrix}.
\ee

\noindent{Here, $\gamma = (\alpha^2 + \beta^2)/2$. To construct a Lorentz covariant quantum field theory of a massless four vector field, the momentum space four vector with the standard momentum $(0,0,k,k)$ have to obey [38]:}

\be
{W^{\mu}_{~\nu}}{e^{\nu}(k,\sigma)} 
= \exp{(- i \sigma \theta(k,W))}{e^{\mu}(k,\sigma)}. 
\ee

\noindent{However, the matrix representing ${S^{\mu}_{~\nu}}$ is not normal 
	$(S^{T}S \neq SS^{T})$. This is also valid for $W = SR$ and the above eigenvalue equation for ${e^{\mu}(k,\sigma)}$ does not exist [66]. Thus, we restrict $W$ to $R_{z}(\theta)$ only. This is consistent with the fact that massless particle states in flat spacetime are characterized by helicity and there are no charges associated with the generators of $S(\alpha, \beta)$ with the charges continuously dependent on $\theta$, [38].} 

We now consider ${e^{\mu}(k,\sigma)}$ that satisfy the following equation instead of Eq.(30):

\be
[R_{z}(\theta)]^{\mu}_{~\nu}{e^{\nu}(k,\sigma)} 
= \exp{(- i \sigma \theta)}{e^{\mu}(k,\sigma)} 
\ee

\noindent{where, $\sigma = 0, \pm{1}$. A convenient choice for $e^{\mu}(k,0; \pm 1)$ is:
	$e^{\mu}(k,\pm 1) = {1\over \sqrt{2}}(1,\pm i,0,0)$ and $e^{\mu}(k,0) = (0,0,1,1)$. We note that for null vectors like $(0,0,k,k)$ in Minkowski space, there can be more than one way to obtain a null vector like: 
	$({{\beta k} \over {\sqrt{1 - \beta^2}}}, 0,k,{{k} \over {\sqrt{1 - \beta^2}}})$.
	Firstly, we can apply a boost $\beta$ along the $x^1$ axis. Alternatively, we can rotate $(0,0,k,k)$ to $(k \beta,0,k \sqrt{1 - \beta^2},k)$ and thereafter apply a boost along the direction of the spatial momentum by $\alpha$ where:
	
	\be
	1 - \beta^2 = {{1 - \alpha} \over {1 + \alpha}}, ~~ \alpha = |\vec{\alpha}|.
	\ee

	\noindent{The above two Lorentz transformations do not lead to the same ${e^{\mu}(\vec{p},\pm 1)}$. This is related to the four -dimensional geometry. However, we can always have: ${k_{\mu}}{e^{\mu}(k,\sigma)} = {p_{\mu}}{e^{\mu}(\vec{p},\sigma)} = 0 = {\partial_{\mu}}{B^{\mu}}(x)$, where ${B^{\mu}}(x)$ is the Fourier transform of ${e^{\mu}(\vec{p},\sigma)}$ given by, [38]:}

	\be
	B^{\mu}(x) = \int {{d^3}p \over \sqrt{2p^0}} {(2\pi)^{-3/2}}
	{\sum_{\sigma}}[{e^{\mu}(\vec{p},\sigma)} {\exp(ip.x)} a(\vec{p},\sigma) + c.c]
	\ee

	\noindent{where, $a(\vec{p},\sigma)$ is an operator. Under a Lorentz transformation $L$, ${B^{\mu}}$ transform as:}
	
	\be
	U(L){B_{\mu}(x)}{U^{-1}}(L) = 
	{L^{\nu}_{~\mu}}{B_{\nu}(L x)}
	\ee

	\noindent{where $U(L)$ are the unitary transformations constructed from a suitable Lagrangian density. To construct a local theory invariant under parity, we can introduce a massless vector field $B_{\nu}$ through the $(1,1)$ type symmetric traceless field: ${{\partial_{(\mu}}{{B}_{\nu)}}}$ (here we have $\partial . B = 0$) or through the ${(1,0) \oplus (0,1)}$ type antisymmetric field ${{\partial_{[\mu}}{{B}_{\nu]}}}$. The latter is used to describe QED to preserve the $U(1)$ internal symmetry present in the corresponding sources [37]. In this context, we note that the Coulomb gauge 
		$\vec{\nabla}. \vec{B} = 0$ together with $B^{0} = 0$ is a non-covariant choice of gauge and a Lorentz transformation on one such field will in general not yield another such field [37]. Things change in QED when we modify the canonical commutators to eliminate the unphysical degrees of freedom. In this case, the right hand side of the above equation is appended by pure gauge terms that ensure Lorentz covariance [68]. The above conclusions are also consistent with the Gupta-Bleuler quantization of the electromagnetic field [37]. ${{\nabla_{(\mu}}{{B}_{\nu)}}}$ is not gauge invariant and $B_{\alpha}$, with the kinetic term obtained using ${{\nabla_{(\mu}}{{B}_{\nu)}}}$, cannot minimally couple with matter fields that possess internal symmetries like $U(1)$ in a theory that preserves such symmetries.
		Thus, the effect of these fields would be purely gravitational to influence the metric
		and they appear only in Einstein's equations. This is similar to the discussions given in Sec.I regarding $(\omega, \psi)$ and can be relevant to explaining dark energy and dark matter. On the other hand, coupling between these fields and matter fields can lead to gauge invariance breaking effects like particle-antiparticle asymmetry. The present section demonstrates that we can use massless higher spin bosons to construct higher spin driven inflation theories [8,9,16] that need not to be gauge theories.}

\section*{V. Conclusion}

To conclude, in this article we have proposed new geometric fields from the gravity sector as possible candidates for inflation and dark energy. Previously, Einstein-Palatini action was used to introduce two massless scalar fields $\omega$ and $\psi$, that give non-metricity. It was found to be more consistent with a quantum theory of gravity. 
The connection is expressed as: $\Gamma^{\alpha}_{~\mu \nu} + C^{\alpha}_{~\mu \nu}(\omega,\psi)$. The Einstein-Palatini formalism is particularly suitable for introducing scalar fields through $C^{\alpha}_{~\mu \nu}$.
We found that for fields relevant to the standard model with the addition of the right handed neutrinos, $C^{\alpha}_{~\mu \nu} (\omega, \psi)$ can at most contribute boundary terms to the matter and radiation field actions and do not appear in the equations of motion. Thus, we do not need to include $C^{\alpha}_{~\mu \nu} (\omega, \psi)$ in the actions and stress tensors of ordinary matter and radiation that are various representations of $SL(2,c)$, through the covariant derivatives. This scheme, developed in Sect.III, can also be required to construct a stable theory with dark energy. The issue of stability is discussed in Sect.IV, where we have performed linear transformation on $\omega$ and $\psi$, to obtain a theory of two decoupled scalars $\Phi$ and $\Psi$. $\Psi$ gives a negative stress tensor. We can introduce potential terms in $\Phi$ and $\Psi$. The negative kinetic term in $\Psi$ prohibits any coupling between itself and other fields, including $\Phi$. We can have a self-interacting theory of $\Psi$. We can couple $\Phi$ to ordinary matter. All these give a versatile model where vacuum expectation values and quantum fluctuations can play important roles. We can construct various models by adding suitable interaction terms. We can generalize the $\Lambda$CDM theory by including $\Phi$ and $\Psi$ to give dynamic and spatially varying dark energy and inflation. $\Lambda$, $\Phi$ and $\Psi$ are similar in many respects. Together, they give the most general theory linear in scalar curvature that is consistent with the Bianchi identity in $\Gamma^{\alpha}_{~\mu \nu}$. They need not to have ordinary matters as their sources and can be non-localized. They contribute positive and negative stress tensors in Einstein's equation that are not coupled to the metric through the gravitational constant. $\Lambda$, $\Phi$ and $\Psi$ give us departures from the classical theory and the global splitting of spacetime into space and time, thus modifying the global structure of spacetime. However, present cosmological observations suggest that such effects are finite but small. Eq.(29) can be used to construct new non-trivial solutions of the vacuum Einstein's equation. Lastly, we can consider a theory that contains more than one $\Phi$ fields but no $\Psi$ field.

We found that we may have to introduce the right-handed neutrinos into the standard model to define the equation of motion and conserved vector currents for neutrinos in curved spacetime. This is related to the existence of required variational derivatives in curved spacetime. The same remains valid for curvilinear coordinates in flat spacetime. The right handed neutrinos have been used before to eliminate anomalies. The axial vector currents for various Dirac fields including the neutrinos can still remain anomalous. Anomalies in axial vector currents are common in flat spacetime. We considered the right-handed neutrinos to be free field singlets of the $SU(2)$ gauge theory in the present article. They can be useful in the dark matter research. They can also be useful to explain neutrino oscillation. 
The neutrinos couple with $\Phi$ and $\Psi$ when we do not include the right handed neutrinos in the action. 
In this case, we can not define conserved neutrino vector currents in curved spacetime even if we use the Levi-Civita connection. The above observations can be important in the early universe.

In this article, we have used $(A,A)$ type representations to introduce higher integral spin fields. We have found that for massless fields, we have to use a one parameter subgroup of $SO(3)$ as the little group in Minkowski space. This allows us to use $(A,A)$ type fields to introduce massless integral spin fields with spin greater than zero. Such fields do not couple with ordinary matter through gauge theories. These fields and the right handed neutrinos can be useful in dark matter, dark energy and higher spin driven inflation models. We can modify the action or use $f(R')$ theories to include higher spin fields. Lastly, we have introduced a phantom scalar using torsion that gives non-metricity. It couples with gauge fields and can violate current conservation laws. Its vacuum effects can be useful in cosmology.

\section*{Acknowledgement}

I am thankful to a few reviewers for some improvements, [24].

\section*{References}

[1] A. A. Starobinisky: A new type of isotropic cosmological models without singularity. Phys. Lett. B {\bf 91}, 99 (1980).

[2] A. H. Guth: Inflationary universe: A possible solution to the horizon and flatness problems. Phys. Rev. D {\bf 23}, 347 (1981).

[3] D.N. Spergel \textit{et al} [WMAP Collaboration]: First year Wilkinson microwave anisotropy probe (WMAP) observations: Determination of cosmological parameters. Astrophys. J. Suppl. {\bf 148}, 175 (2003). 

[4] D.N. Spergel \textit{et al} [WMAP Collaboration]: Three-Year Wilkinson microwave anisotropy probe (WMAP) observations: implications for cosmology. Astrophys. J. Suppl. {\bf 170}, 377 (2007).

[5] E. Komatsu \textit{et al} [WMAP Collaboration]: Five-Year Wilkinson microwave anisotropy probe observations: cosmological interpretation. Astrophys. J. Suppl. {\bf 180}, 330 (2009).

[6] L. Amendola and S. Tsujikawa: Dark Energy. Cambridge University Press, Cambridge, 2010.

[7] A. D. Linde: Axions in inflationary cosmology, Phys. Lett. B \textbf{259}, 38 (1991).

[8] N. Kaloper: Lorentz Chern-Simons terms in Bianchi cosmologies and the cosmic no hair conjecture. Phys. Rev. D {\bf 44}, 2380 (1991).

[9] A. Golovnev, M. Mukhanov and V. Vanchurin: Vector inflation. JCAP {\bf 0806:009}, 2008.

[10] F. Zwicky,: Die Rotverschieb ung von extragalaktischen Nebeln. Helv. Phys. Acta {\bf 6}, 110 (1933).

[11] V. Rubin, N. Thonnard and W.K. Ford, Jr., Rotational properties of 21 Sc galaxies with a large range of luminosities and radii from NGC 4605 (R = 4 kpc) to UGC 2885 (R = 122 kpc)". The Astrophysical Journal. 238: 471 (1980).

[12] Bertone, G. Bertone, D. Hooper and J. Silk, "Particle dark matter: Evidence, candidates and constraints". Physics Reports. 405 (5–6): 279–390 (2005).

[13] S. Kobayashi and K. Nomiju: Foundation of Differential Geometry Vol I. John Wiley and Sons, Inc., 1991.

[14] S. Weinberg: The cosmological constant problem. Rev. Mod. Phys. {\bf 61}, 1 (1989).

[15] V. Sahni and A. A. Starobinsky, Int. J. Mod. Phys. D 9, 373 (2000).

[16] D. Blais and D. Polarski, Phys. Rev. D 70, 084008 (2004).

[17] A. Y. Kamenshchik, U. Moschella and V. Pasquier: An alternative to quintessence. Phys. Lett. B {\bf 511}, 265 (2001).

[18] S. Capozzillo: Curvature quintessence. Int. J. Mod. Phys. D {\bf 11}, 483 (2002).

[19]  A. De Felice and S. Tsujikawa: f(R) theories. Living Reviews in Relativity. {\bf 13} (2010).

[20] R. Gannouji, D. Polarski, A. ranquet and A. A. Starobinski: Scalar-tensor models of normal and phantom dark energy. \textit{JCAP} \textbf{0609}, 016 (2006).

[21] G. R. Dvali, G. Gabadadze and M. Porrati: 4-D gravity on a brane in 5-D Minkowski space. Phys. Lett. B {\bf 485}, 208 (2000). 

[22] V. Sahni and Y. Shtanov: Braneworld models of dark energy. JCAP {\bf 311}, 014 (2003).

[23] K. Ghosh: Non-metric fields from quantum gravity. Proceedings of the MG15 Meeting on General Relativity, Part A, University of Rome “La Sapienza”, Italy, 1 – 7 July 2018, https://doi.org/10.1142/12843; Edited By: Elia S Battistelli, Robert T Jantzen and Remo Ruffini; https://hal.science/hal-02105422v4

[24] K. Ghosh: Affine connections in quantum gravity and new scalar fields. Physics of the Dark Universe, \textbf{26} (2019) 100403; arXiv:1910.12633v5 [gr-qc].
 
[25] K. Ghosh: Affine Connection, Quantum theory and New Fields, to be published in Quantum Studies: Mathematics and Foundations. K. Ghosh: https://hal.science/hal-04086265

[26] J. G. Hocking and G. S. Young: Topology. Dover Publications, Inc., New York, 1961.

[27] D. Lovelock and H. Rund: Tensors, Differential Forms, and Variational Principals. Dover Publications, Inc., New York, 1989.

[28] L. D. Landau and E. M. Lifshitz: The Classical Theory of Fields. Butterworth-Heinenann, Oxford, 1998.

[29] R. M. Wald: General Relativity. The University of Chicago Press, Chicago and London, 1984.

[30] C. W. Misner, K. S. Thorne and J. A. Wheeler: Gravitation. W.H. Freeman and company, New York, 1970. 

[31] F. W. Hehl \textit{et al}: General relativity with spin and torsion: Foundations and prospects. Rev. Mod. Phys. {\bf 48} (1976) 3641.

[32] A. R. Liddle and D. H. Lyth: Cosmological Inflation and Large Scale Structures.
Cambridge University Press, 2000.

[33] S. Dodelson and F. Schmidt, Modern Cosmology. Academic Press, 2021. 

[35] J. V. Narlikar: Mini-bangs in cosmology and astrophysics. Pramana {\bf 2}(3): 158–170 (1974).

[36] L. H. Ford and T. A. Roman: Negative Energy, wormholes and warp Drive. Scientific American {\bf 282} 46 (2000).

[37] C. Itzykson and J. B. Zuber: Quantum Field Theory. Dover Publications, Inc. Mineola, 2005. 

[38] S. Weinberg: The Quantum Theory of Fields, Vol.\textbf{I}. Cambridge University
Press, Cambridge, 1996.

[39] N. D. Birrell and P. C. W. Davies: Quantum Fields in Curved Spaces. Cambridge University Press, 1994.

[40] S. Weinberg: Gravitation and Cosmology. John Wiley $\&$ Sons, 2004.

[41] T. Yanagida: Horizontal symmetry and mass of the t quark. Phys. Rev. D \textbf{20}, (1979), 2986. 

[42] T. Yanagida: Horizontal Symmetry and masses of neutrinos. Proc. Theo. Phys. \textbf{64}, No.3, (1980), 1103. 

[43] M. E. Preskin and D. V. Schroeder: An Introduction to Quantum Field Theory. Addison-Wesley Publishing Company, 1995.

[44] S. Weinberg: The Quantum Theory of Fields, Vol.II. Cambridge University
Press, Cambridge, 1996.

[45] P. Ramond: Field Theory: A Modern Primer. Addison-Wesley Publishing Company, 1990.

[46]  M. Gasperini and G. Veneziano: The Pre-Big bang scenario in string cosmology. Phys. Rept. \textbf{373}, 1 (2003).

[47] R. Gilmore: Lie Groups, Lie Algebras, and Some of Their Applications. Dover Publications, Inc. Mineola, 2002.

[48] S. W. Hawking and G. F. R. Ellis: The Large Scale Structure of Space-Time. Cambridge University Press, 1973.

[49] H. Stephani, et al.: Exact solutions of Einstein's field equations. Cambridge University Press, 2003.

[50] S. Dodelson and F. Schmidt, Modern Cosmology. Academic Press, 2021. 

[51] R. R. Caldwell, M. Kamionkowski, and N. N. Weinberg: Phantom Energy and Cosmic Doomsday, Phys. Rev. Lett. \textbf{91} (2003), 071301.

[52] S. M. Carroll, M. Hoffman and M. Trodden: Can the dark energy equation-of-state parameter $w$ be less than $-1$? Phys. Rev.D \textbf{70} (2003), 023509.

[53] P. Singh, M. Sami and N. Dadhich: Cosmological dynamics of phantom field. Phys. Rev.D \textbf{68} (2003), 023522.

[54] J. M. Cline, S. Jeon, and G.D. Moore, The phantom menaced: Constraints on low energy effective ghosts, Phys. Rev. D \textbf{70} (2004), 043543.

[55] R. R. Caldwell, M. Kamionkowski, and N. N. Weinberg: Phantom Energy and Cosmic Doomsday, Phys. Rev. Lett. \textbf{91} (2003), 071301.

[56] H. Bondi and T. Gold: The steady-state theory of the expanding universe. Mon. Not. Roy. Ast. Soc. \textbf{108} (1948), 252-70.

[57] F. Hoyle: A new model for the expanding universe. Mon. Not. Roy. Ast. Soc. \textbf{108} (1948), 372-82.

[58] F. A. E. Pirani: On the energy-momentum tensor and the creation of matter in relativistic cosmology, Proc. Roy. Soc. Lond. A \textbf{228} (1955), 455-62. 

[59] F. Hoyle and J.V. Narlikar: A new theory of gravitation. Proc. Roy. Soc. Lond. A \textbf{277} (1964), 1-23.

[60] J. Bub: Interpreting the Quantum World. Cambridge University Press, 1997.

[61] A. Fine: The Shaky Game. The University of Chicago Press, 1996.

[62] A. Aspect: Closing the door on Einstein and Bohr's quantum debate. Physics \textbf{8}, 123 (2015).

[63] H. B. Sandvik, J. D. Barrow and J. Magueijo, A simple varying-alpha cosmology. Phys. Rev. Lett. \textbf{88} (2002), 031302.  

[64] T. Chiba and K. Kohri, Quintessence cosmology and varying alpha.  Prog. Theor. Phys. \textbf{107} (2002), 631. 

[65] D. Parkinson, B. A. Bassett and J. D. Barrow, Mapping the dark energy with time-varying alpha. Phys. Lett. B \textbf{578} (2004), 235.

[66] R. T. Hammond: Torsion gravity. Rep. Prog. Phys. {\bf 65}, 599 (2002).

[67] H. Joos: Fortschr. Phys. 10\textbf{10}, 65 (1962).

[68] J. D. Bjorken and S. D. Drell: Relativistic Quantum Fields. McGraw-Hill Book Company, 1995.

[69] C. G. Cullen: Matrices and Linear Transformations. Dover Publications, Inc., New York, 1972.





\end{document}